# Inhomogeneously doped thermoelectric nanomaterials

T. E. Humphrey and H. Linke
[1]Engineering Physics, University of Wollongong, Wollongong 2522, Australia.
[2]Centre of Excellence for Advanced Silicon Photovoltaics and Photonics, University of New South Wales, N.S.W 2052, Australia. Email: tammy.humphrey@unsw.edu.au; Web: www.humphrey.id.au
[3]Materials Science Institute and Physics Department, University of Oregon, Eugene OR 97403-1274, U.S.A.

**Abstract**

Inhomogeneously doped thermoelectric nanomaterials with a delta-function electronic density of states can operate with Carnot efficiency in the absence of phonon heat leaks. Here we self-consistently calculate the efficiency and power from open-circuit to short-circuit of a simple model of a thermoelectric nanomaterial with a narrow peak in the electronic density of states and finite lattice thermal conductivity, comparing the results for inhomogeneous and homogeneous doping. For power generation between 800K and 300K, we find that not only does inhomogeneous doping increase the maximum efficiency by 10%, but it also increases the maximum power by up to 60%.

**Background**

High efficiency solid-state power generators and refrigerators have enormous potential for applications in, among many others, the automotive, microelectronics and refrigeration industries. A recent breakthrough has been the development of *nanostructured* thermoelectric materials [1-3] with remarkably high figures of merit, $ZT = T\sigma S^2/\kappa$ (where $T$ is the temperature, $\sigma$ the electrical conductivity, $S$ the Seebeck coefficient and $\kappa = \kappa_{ph} + \kappa_{el}$ is the sum of the lattice and electronic contributions to the thermal conductivity of the material) that are thought to result from a combination of two distinct effects [4-5]. Firstly, it is known that phononic heat conduction is reduced in materials with a high interface density [6-9]. Secondly, quantum confinement effects can produce sharp peaks in the electronic density of states (DOS) [10]. While the underlying physical mechanism has not been clear [5], delta-like DOS have been found to result in (i) an improvement in the thermopower, $S$, without a corresponding reduction in electrical conductivity, $\sigma$, [2,10-11], (ii) an electronic contribution to the thermal conductivity, $\kappa_{el}$, equal to zero [12] which leads to (iii) a theoretical maximum in $ZT$ [12].

In another paper [13], we discuss the fundamental thermodynamics responsible for this second group of effects and derive an analytic expression for the spatial variation in electrochemical potential needed to achieve reversible electron transport in a thermoelectric material with a delta-function electronic density of states (DOS) to be [13]:

$$\mu_0(\mathbf{r}) = E_0 - eS_0 T(\mathbf{r}) \quad (1)$$

where $\mu_0(\mathbf{r})$ is the electrochemical potential across the n- or p-type leg of a thermoelectric nanomaterial with a delta-like DOS centered on the energy $E_0$, $S_0$ is the theoretical maximum, spatially invariant Seebeck coefficient [13] and $T(\mathbf{r})$ is the temperature profile across the nanomaterial. Here we build directly upon this work by quantifying in detail the improvement in the efficiency and power gained by having an electrochemical potential which varies according to Eq. 1. in nanomaterials with delta-like DOS.

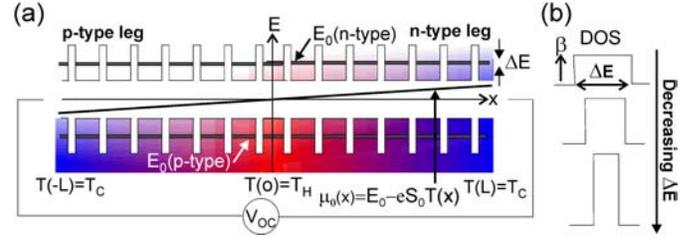

**Figure 1(a):** Simplified schematic of the bandstructure of a thermoelectric quantum dot superlattice (QDSL) or superlattice nanowire (SLNW) with a single miniband in the conduction and valence bands yielding a delta-like DOS with width $\Delta E$ for electrons (holes) located at energy $E_0$ ($-E_0$). The doping level varies across the material to ensure that $\mu(x)$ satisfies Eq. 1. **(b)**: Schematic of the scaling procedure for $\beta$ (see Eq. 2), showing that as $\Delta E$ decreases, the magnitude of $\beta$ is correspondingly increased to maintain a constant number of available states for electrons.

**Model**

Figure 1a shows a simplified band-structure schematic of a thermoelectric material such as a quantum dot superlattice (QDSL) or superlattice nanowire (SLNW) in which the formation of narrow, well-separated minibands due to quantum confinement effects results in a delta-like electronic DOS. We seek to obtain the spatial dependence of $T(x)$, $S(x), \sigma(x)$, $\kappa_{el}(x)$ and $ZT(x)$, and thereby power and efficiency curves from open-circuit to short circuit conditions for the nanomaterial in Fig. 1a as a function of the width of the DOS peak for both a constant electrochemical potential (which we will denote as 'homogenous doping') and one which varies according to Eq. 1 (denoted as 'inhomogeneous doping').

At a particular point in the material $S$, $\sigma$, $\kappa_{el}$ and $ZT$ may be calculated from the Boltzmann transport equation under the relaxation-time approximation [14-15], and can all be expressed as a function of the integral [14-15]

$$K_\alpha = \int \beta(E)(E-\mu)^\alpha \left(-\frac{df}{dE}\right) dE \quad (2)$$

where $\beta(E) = D(E)\tau(E)v(E)$, $\tau(E)$ is the electron relaxation time, $v(E)$ is the electron group velocity, $D(E)$ is the DOS and where $\alpha$ = 0, 1 or 2. Then $\sigma = e^2 K_0$, $S = -K_1/(eTK_0)$, $\kappa_{el} = (K_2 - K_1^2/K_0)/T$. For simplicity and physical transparency in the results, we assume that $\beta(E) = \beta$ is constant over the energy range $\Delta E$, a reasonable assumption for the small values of $\Delta E$ in which we are primarily interested. To isolate effects upon $ZT$ due to the width of the DOS from effects due to changing the overall *number* of available states for electrons, we vary $\beta$ with $\Delta E$ (see Fig. 1b) such that $\beta(\Delta E) = 5\times10^5/e^2K_0(\Delta E,\mu_H=E_0)$. This means that for all values of $\Delta E$, $\sigma = 5\times10^5$ $\Omega^{-1}m^{-1}$ if the miniband is centered on the Fermi energy, $E_0 = \mu(0)$.

**Numerical Technique**

Following the procedure outlined by Mahan [16] in his paper analyzing inhomogeneous doping in bulk thermoelectric materials, we note that Domenicali's equation for energy balance and the equation for heat flow can be used to obtain the two differential equations:

$$\frac{dT(x)}{dx} = \frac{IT(x)S(x) - J_q(x)}{\kappa_{el}(x) + \kappa_{ph}} \quad (3)$$

and

$$\frac{dJ_q(x)}{dx} = \frac{I^2[1 + ZT(x)]}{\sigma(x)} - \frac{IS(x)J_q(x)}{\kappa_{el}(x) + \kappa_{ph}} \quad (4)$$

where $I$ is the electrical current density and $J_q(x)$ the heat current density at a point $x$ in the material. The numerical procedure is to first specify that $T(0) = T_H$ (where $T_H = 800$K) then make an initial guess for $J_q(0)$, feed this into equations 3 and 4, iterate across the material from $x = 0$ to $x = L$, then to compare the temperature $T(L)$ so obtained with the desired temperature, $T_C = 300$K. The initial guess for $J_q(0)$ is then adjusted and the procedure repeated until $T(L) = T_C$ is obtained to a suitable tolerance (we used $T(L) = T_C \pm 0.001$K). The thermoelectric parameters $S(x)$, $\kappa_{el}(x)$, $\sigma(x)$ and $ZT(x)$ are calculated for each point along the $x$-axis using Eq. (2). For inhomogeneous doping we used $\mu_{IH}(x) = E_C - eS(x)T(x)$, for homogeneous doping $\mu_H = \mu_{IH}(L/2)$. For a particular combination of $\kappa_{ph}$ and $\Delta E$, the above procedure is repeated for different values of $E_0 - \mu(0)$ to find the value which maximizes $ZT(0)$ at open circuit.

The power of the thermoelectric material is given by $IV$, and the efficiency by:

$$\eta = IV/Q_H \quad (5)$$

where

$$V = \int_0^L \left\{ S(x)\frac{dT(x)}{dx} + \frac{I}{\sigma(x)} \right\} dx \quad (6).$$

The above calculation is then repeated for different values of current from zero to short-circuit current, to obtain entire power-efficiency curves for both homogenous and inhomogenous doping, for different values of $\kappa_{ph}$ and $\Delta E$.

**Results**

Figure 2 shows $S(x)$, $\kappa_{el}(x)$, $\sigma(x)$ and $ZT(x)$ for $\kappa_{ph} = 0.5$ Wm$^{-1}$K$^{-1}$ and $I = -0.01$ Am$^{-2}$ (essentially open-circuit, negative current implies electrons flow from hot to cold) as a function of $T(x)$ for four different widths of the DOS, $\Delta E = 10, 60, 100$, and $250$meV (where arrows indicate the direction of decreasing $\Delta E$), for both inhomogeneous (red solid lines) and homogeneous (blue dotted lines) doping. Figure 2d illustrates one of the main results of the paper, that $ZT(x)$ increases with both decreasing $\Delta E$ and with a change from homogeneous to inhomogeneous doping according to Eq. 1. It can also be seen that $S$ for inhomogeneous doping tends to become constant across the material as $\Delta E$ decreases, as does $ZT$. This occurs as $S \rightarrow S_0$, the spatially constant theoretical maximum Seebeck coefficient derived in [13]. This does not occur for homogeneous doping with decreasing $\Delta E$.

As discussed in [13], the value of $[E_0 - \mu(0)]$ which maximizes $ZT(0)$ decreases as $\Delta E$ decreases to approach $2.4kT$, the value derived by Mahan and Sofo for a $\delta$-function DOS [12]. This decrease in $[E_0 - \mu(0)]$ for smaller $\Delta E$ has two effects, firstly it means that value of the $S$ at which $ZT$ is maximised (shown in Fig. 2a) becomes smaller in magnitude, and that $\sigma$ (shown in Fig. 2b) increases. Although $\sigma$ in Fig 2b increases by a factor of four between $\Delta E = 250$meV and $\Delta E = 10$meV, $\kappa_{el}$ actually *decreases* by almost two orders of magnitude, illustrating the fact that the Wiedemann-Franz law [14] is not applicable to materials with delta-like DOS [13,16].

Results for the power and efficiency relative to the Carnot limit of both inhomogeneous (red solid lines) and homogeneous doping (blue dotted lines) for the four different DOS widths are shown in Fig 3a to 3c for $\kappa_{ph} = 0.2, 0.5$ and $1.0$ Wm$^{-1}$K$^{-1}$ respectively. Current increases and voltage decreases between open and short circuit conditions (see Fig. 3d) as the loops are traversed clockwise from the point where both power and efficiency are zero. As in Fig 2, arrows indicate increasing $\Delta E$. The first observation that can be made about the results shown in Fig 3a - c is that the smaller the value of $\Delta E$ the higher the power and efficiency for both homogeneous and inhomogeneous doping (note that in our model we have assumed the number of states available for electrons is independent of $\Delta E$). This supports Mahan and Sofo's result [12] that a delta-like DOS is optimum for efficiency, and additionally shows that device power also has the potential to benefit substantially from nanostructuring of thermoelectric materials.

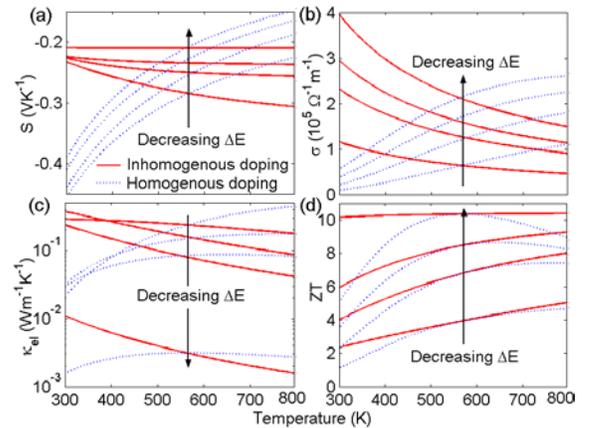

**Figure 2: (a-d)** $\Delta E = 10, 60, 100,$ and $250$meV, where arrows indicate the direction of decreasing $\Delta E$, for both inhomogeneous (red solid lines) and homogeneous (blue dotted lines) doping, for $\kappa_{ph} = 0.5$ Wm$^{-1}$K$^{-1}$ and $I = -0.01$ Am$^{-2}$. (a) Seebeck coefficient across the material shown in Fig 1a. (b) Electrical conductivity. (c) Thermal conductivity due to electrons. (d) Dimensionless figure of merit.

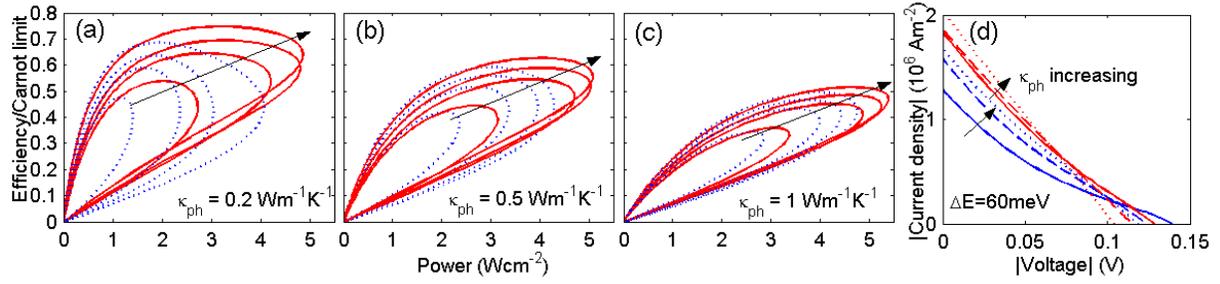

**Figure 3: (a)-(c)** 'Loop' plots of efficiency relative to the Carnot limit versus power for inhomogeneous doping (red solid lines) and homogeneous doping (blue dotted lines) for the same values of $\Delta E$ as in Fig 2, for $\kappa_{ph} = 0.2, 0.5$ and $1$ Wm$^{-1}$K$^{-1}$. The arrows indicate decreasing $\Delta E$. **(d)** Current-voltage curves for the $\Delta E = 60$ loops in Fig 3a-c. Inhomogeneous doping results are shown in red, homogeneous in blue for $\kappa_{ph} = 0.2$ (dotted lines), $0.5$ (dashed lines) and $1.0$ Wm$^{-1}$K$^{-1}$ (solid lines). Arrows indicate increasing $\kappa_{el}$ for both homogeneous and inhomogeneous doping.

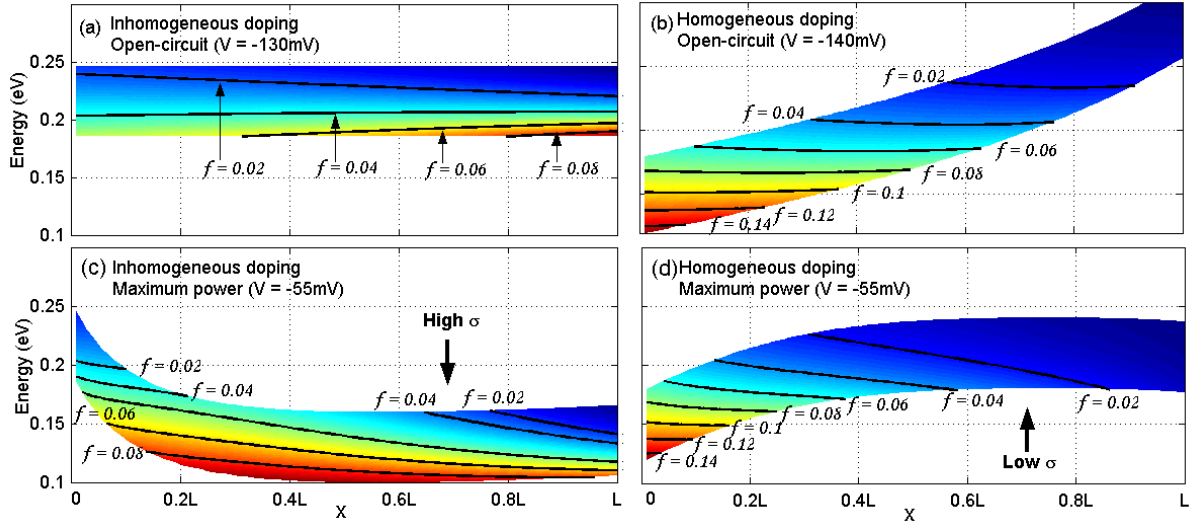

**Figure 4:** Contour plots of the occupation of available states as a function of energy and position along the *x*-axis in a 60 meV wide miniband in the n-type leg of a thermoelectric nanomaterial such as that illustrated in Fig 1a with $\kappa_{ph} = 0.2$ Wm$^{-1}$K$^{-1}$ for: **(a)** inhomogeneous doping according to Eq. 1 at open circuit, **(b)** homogeneous doping at open circuit, **(c)** inhomogeneous doping at maximum power and **(d)** homogeneous doping at maximum power. Lines of constant occupation of states are marked, and separated by a difference in occupation of 0.02. Energies are measured from the center of the bandgap at $x = 0$. As the magnitude of the applied voltage is reduced between (a) and (c) and between (b) and (d), the energy of states in the miniband at $x = L$ accordingly shifts to lower energies.

Secondly, it can be seen that the maximum power increases as $\kappa_{ph}$ increases. While this result may appear counterintuitive, it is due to the fact that this relatively simple model assumes hot and cold reservoirs of infinite heat capacity, so that the higher currents (shown in Fig 3d) needed to obtain open-circuit conditions for higher $\kappa_{ph}$ can be drawn without lowering the temperature of the hot reservoir or raising that of the cold reservoir. In practice there is a limit on the current that can be drawn while maintaining a particular temperature gradient across a material.

Finally, our main result: that inhomogeneous doping to achieve the electrochemical potential given by Eq. 1 gives a higher power and efficiency for all values of $\Delta E$ and $\kappa_{ph}$. Our results for relative efficiency improvement (of the order of 10%) are similar to those obtained by Mahan for inhomogeneously doped bulk materials [16]. What is most remarkable however, is the very large relative increase (up to 60%) in the maximum power. In Fig. 3d we show the current-voltage curves corresponding to the six 'loops' in Fig 3a-c for which $\Delta E = 60$ meV. For $\kappa_{ph} = 0.2$ Wm$^{-1}$K$^{-1}$ and $V \approx 55$ mV (the voltage where maximum power is obtained) the current flowing in the inhomogeneously doped material (red solid line) is much larger than for the homogeneously doped material, directly leading to a higher power in the former.

## Discussion

While we have previously shown that inhomogeneous doping according to Eq. 1 can lead to Carnot efficiency in thermoelectric nanomaterials with a delta-function DOS in the case where the phonon conductivity tends to zero [13], it is clear from the above results that it is not just the efficiency, but also the maximum power that is improved by inhomogeneous doping. To provide a physical explanation for why inhomogeneous doping according to Eq. 1 results in a higher current and therefore higher maximum power, in Fig 4 we show a contour plot of the Fermi occupation function

$$f(E,x) = \left[\exp\left(\frac{E-\mu(x)}{kT(x)}\right)+1\right]^{-1} \quad (7)$$

as a function of position along the *x*-axis and energy of available states within the miniband in the n-type leg of a nanomaterial such as that shown in Fig. 1a at voltages corresponding to open-circuit and maximum power conditions for both inhomogeneous and homogeneous doping. The Boltzmann transport equation can be used to obtain an expression for the current (closely related to Eq. 2) as [17]:

$$I(x) = e\int v_x(E)^2 \tau(E) D(E) \frac{df(E,x)}{dx} dE \quad (8)$$

where $v(E)$, $\tau(E)$ and $D(E)$ are defined as for Eq. 2. The current flowing at a particular energy is thus proportional to the gradient of the Fermi distribution. The larger $df(E,x)/dx$, the higher the energy resolved current.

Fig. 4a shows the Fermi occupation of available states in the miniband for open circuit and inhomogeneous doping. It can be seen that at the highest energies in the miniband there is a net flow of electrons from the hot ($x=0$) to the cold ($x=L$) extremes of the material, while at low energies there is a compensating flow from cold to hot such that the net current is zero (open-circuit). This two-way electron flow for finite $\Delta E$ results in a net heat transfer but no work, such that the efficiency and power at open-circuit is zero (as shown in the loop plots in Fig 3). The fact that electrons flowing in opposite directions in thermoelectric material results in an efficiency lower than the Carnot limit was originally noted by Littman and Davidson [18] leading them to conclude that 'no thermoelectric device can ever reach Carnot efficiency' [18], a conclusion which has remained in favour [19]. If, however, the width of the miniband is infinitesimal ($\Delta E = 0$) and is centered at the energy where the opposing effects of the temperature and electrochemical potential difference exactly cancel (in other words, where the Fermi occupation function is constant right across the material) then the energy resolved current is zero, and it can be shown that Carnot efficiency may be obtained in the limit of zero phonon heat leaks [13]. It is important to note that in if the material is homogeneously doped then, as shown in Fig. 4b for open circuit conditions, there is no one energy at which the Fermi occupation function is constant right across the material, and Carnot efficiency cannot be obtained even in the limit that the density of states is a delta-function ($\Delta E = 0$).

Fig. 4c and 4d show the occupation of states in the miniband for inhomogeneous and homogeneous doping at a voltage of $V = 0.55$mV, corresponding to maximum power conditions. We note that the effect of inhomogeneous doping is to increase the average gradient of the Fermi distribution across the material at energies within the miniband. This can be seen qualitatively by noting the spacing of the lines of constant occupation (which indicate the gradient of the Fermi distribution) is closer from $0.2L$ to $L$ in the miniband shown in Fig. 4c than for the same region in the miniband shown in Fig. 4d. An alternate but equivalent physical explanation is to note that the average value of $df(E,x)/dE$, proportional to the energy resolved conductivity (via Eq. 2), is increased through inhomogeneous doping, as the average occupation of states in the miniband is higher in Fig. 4c than that in Fig. 4d.

In summary, we have analyzed the effects of inhomogeneous doping in thermoelectric nanomaterials, finding that improvements in the efficiency of 10%, and in the maximum power of up to 60% may be obtained in our relatively simple model. It is important to note that this improvement has been obtained in a system optimized for efficiency rather than power. It has recently been shown the energy spectrum of transmitted electrons necessary to achieve maximum power in nanostructured thermionic devices [20], is quite different to that required to achieve maximum efficiency [21]. Now that the electronic DOS necessary to achieve maximum efficiency in thermoelectric nanomaterials is known [12, 13], an interesting question is whether this design is also optimum for achieving maximum power.

**Acknowledgments**

TH acknowledges the support of the Australian Research Council and a grant from the NSW Sustainable energy research and development fund. HL acknowledges the support of a CAREER grant from the National Science Foundation.

**References**

1. Ventkatasubramanian, R., Siivola, E., Colpitts, T. and O'Quinn, B., *Nature*, Vol. 412 (2001) pp. 59-602.
2. Harman, T. C., Taylor, P. J., Walsh, M. P. and LaForge, B. E., *Science*, Vol. 297 (2002), pp. 2229-2232.
3. Hsu, K. F., *et al.*, *Science*, Vol. 303 (2004), pp. 818-821.
4. Majumdar, A., *Science*, Vol. 303 (2004), pp. 777-778.
5. Chen, G. *et al.*, *Int. Mat. Rev.*, Vol. 48, No. 1 (2003), pp. 1-22.
6. Hicks, L. D. and Dresselhaus, M. S., *Phys. Rev. B*, Vol. 47, No. 19 (1993), pp. 12727-12731.
7. Hicks, L. D. and Dresselhaus, M. S., *Phys. Rev. B*, Vol. 47, No. 24 (1993), pp. 16631-16634.
8. Cahill, D. G., et al., *J. Appl. Phys.*, Vol. 93, No. 2 (2003), pp. 793-818.
9. Costescu, R. M., et al. *Science*, Vol. 303 (2004), pp. 989-990.
10. Lin, Yu-Ming and Dresselhaus, M. S., *Phys. Rev. B*, Vol. 68 (2003), Art. no. 075304.
11. Balandin, A. A. and Lazarenkova, O. L., *Appl. Phys. Lett.*, Vol. 82, No. 3 (2003), pp. 415-417.
12. Mahan, G. D. and Sofo, J. O., *Proc. Nat. Acad. Sci.*, Vol 93 (1996), pp. 7436-7439.
13. Humphrey, T. E. and Linke, H., submitted to *Phys. Rev. Lett.* cond-mat/0407509 (2004).
14. Ashcroft, N. W., and Mermin, N. D., <u>Solid State Physics</u>, Saunders College Publishing (1976), pp. 253-255.
15. Nolas, G. S., Sharp, J. and Goldsmid, H. J., <u>Thermoelectrics: Basic Principles and New Materials Developments</u>, Springer (Berlin, 2001), pp. 36-37.
16. Mahan, G. D., *J. Appl. Phys.*, Vol. 70 (1991), pp. 4551-4554.

17. Kittel, C., <u>Introduction to Solid State Physics</u>, John Wiley & Sons, Inc. (7th Ed. 1996), pp. 648-649.
18. Littman, H. and Davidson, B., *J. Appl. Phys.*, Vol. 32, No. 2 (1961), pp. 217-219.
19. Mahan, G. D., *Sol. State Phys.,* Vol. 51 (1998) pp. 81-157.
20. Humphrey, T.E. and Linke, H., (2004) [cond-mat/0401377](cond-mat/0401377)
21. Humphrey, T. E., Newbury, R., Taylor, R. P. and Linke, H., *Phys. Rev. Lett.*, Vol. 89 (2002), Art. no. 116801.